%% file: main.tex
\documentclass{article}

\input{preamble}

\tcbuselibrary{skins}
\usepackage{titlesec}
\usepackage[labelfont=bf, font=small, labelsep=period]{caption}
\usepackage[absolute,overlay]{textpos}
\usepackage{microtype} 
\usepackage{listings}

\newcommand{\method}{EvoCap}

\definecolor{DeepIndigo}{RGB}{10, 20, 100}
\definecolor{ElectricCyan}{RGB}{0, 220, 255}
\definecolor{NeonYellow}{RGB}{255, 230, 0}
\definecolor{VividRed}{RGB}{220, 20, 10}
\definecolor{DarkCrimson}{RGB}{110, 0, 5}
\definecolor{caspianbg}{RGB}{248, 250, 255} 

\hypersetup{
    colorlinks=true,
    linkcolor=DeepIndigo,
    citecolor=DeepIndigo,
    urlcolor=ElectricCyan
}

\titleformat{\section}
  {\normalfont\Large\bfseries\color{DeepIndigo}}{\thesection}{1em}{}
\titleformat{\subsection}
  {\normalfont\large\bfseries\color{DeepIndigo!80}}{\thesubsection}{1em}{}

\begin{document}

\begin{center}
    \vspace*{-1.2cm}
    \begin{minipage}[c]{0.25\linewidth}
        \includegraphics[height=0.85cm]{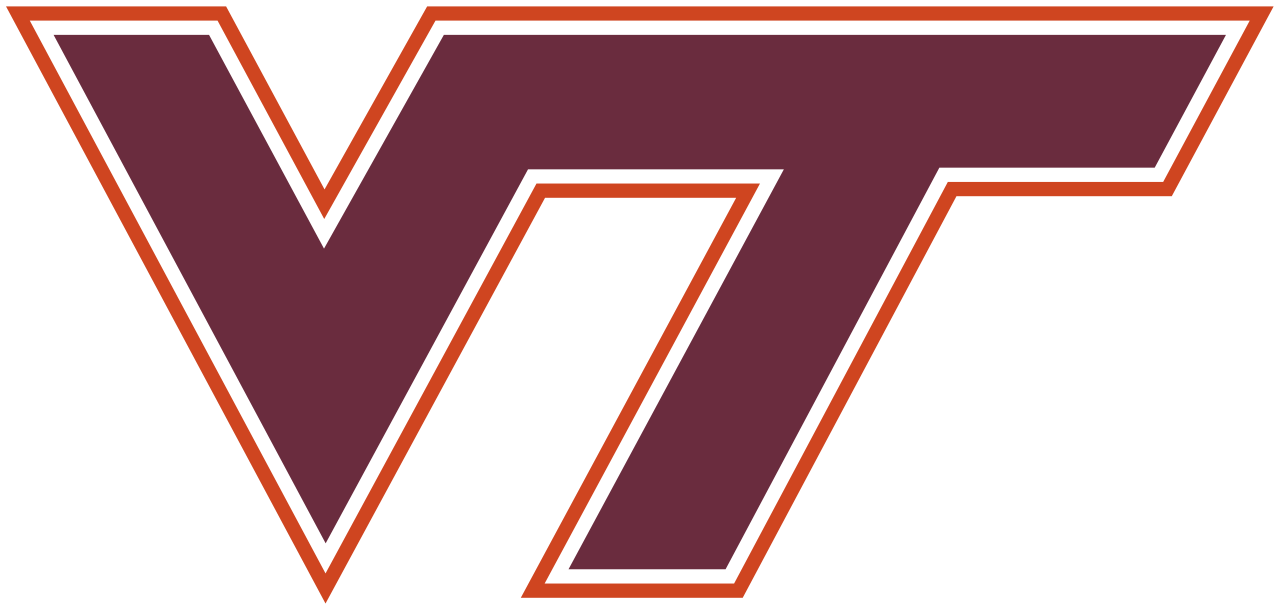}
    \end{minipage}%
    \begin{minipage}[c]{0.5\linewidth}
        \centering
        \footnotesize\textcolor{gray!70}{Preprint \\ \today }
    \end{minipage}%
    \begin{minipage}[c]{0.25\linewidth}
        \hfill 
    \end{minipage}
    
    \vspace{0.25cm}
    {\color{DeepIndigo!20}\rule{\linewidth}{0.6pt}}
    
    \vspace{0.4cm}
\end{center}

\begin{center}
{\LARGE \bfseries
Beyond Tier Labels: Role- and Deployment-Dependent Model Substitution in Multi-Call LLM Workflows\\
\par}

\vspace{0.25cm}

{\large
Renxiang Wang$^{1}$ \quad
Jiaming Cui$^{1}$ \\}
\vspace{0.05cm}
{$^{1}$\texttt{Virginia Tech, Blacksburg, VA}\\}
{\small
\texttt{\{renxiangwang, jiamingcui\}@vt.edu}\\}

\vspace{0.15cm}


\vspace{0.5cm}
\end{center}

\renewenvironment{abstract}{\noindent\ignorespaces}{\par}

\begin{center}
\begin{tcolorbox}[
    enhanced,
    colback=caspianbg,
    colframe=gray!5,
    arc=0mm,
    outer arc=0mm,
    width=\linewidth,
    left=4mm,
    right=4mm,
    top=3mm,
    bottom=3mm,
    boxrule=0pt,
    borderline west={2.5pt}{0pt}{DeepIndigo!90!ElectricCyan}, 
    before skip=10pt,
    after skip=20pt
]
Large multi-call LLM systems pose a scientific problem that query-level routing does not capture: the value of a model depends on where it enters a dependent computation and on the deployment that surrounds that call. Existing routers typically decide \emph{where} to spend a stronger model while treating the benefit of the substitution itself as known. We separate these two decisions through a predicate-action factorization and evaluate it in controlled solve-merge-verify workflows spanning 8-64 solve calls and four three-tier model ladders. The resulting evidence reveals a consistent principle beneath apparently conflicting outcomes. On numeric frequency counting, all-strong reduces RMSE from 4.818 to 1.538 in the Mixed Qwen/GPT ladder, whereas the average Qwen-only ordering reverses. Input-matched interventions further show that the same medium-to-strong action has sharply different value across roles and scales. A semantic task-and-contract shift reverses the Mixed ordering again, while allocation ablations distinguish useful sparse placement from under-coverage and indiscriminate escalation. Together, these results establish model substitution as a deployment-conditioned action rather than a property implied by a tier label, and they provide a practical sequence for large-scale workflow routing: calibrate the action, resolve its role-conditioned effect, and then optimize its placement.
\end{tcolorbox}
\end{center}

\vspace{0.1cm}

\section{Introduction}\label{sec:intro}

Large-scale LLM agent systems distribute a task across role-specialized calls, enabling a single workflow to combine debate, software production, and graph-shaped collaboration \citep{li2023camel,hong2024metagpt,qian2024chatdev}. Scaling, however, changes not only the computational budget but also the scientific object. Each additional call produces an intermediate representation that downstream calls must interpret, reconcile, or verify. Consequently, system quality is mediated by the surrounding graph. This dependence helps explain why adding agents alone does not guarantee a better answer \citep{du2024debate,li2024moreagents,qian2025scaling}, even though aggregation can improve final outputs \citep{wang2025moa,jiang2023llmblender}. To make these gains practical, resource-aware methods reduce the associated cost by pruning agents, communication, or budgeted interactions \citep{chen2024optima,wang2025agentdropout,yang2026bamas}. Yet a more fundamental scientific question remains unresolved: what constitutes a beneficial model intervention inside a dependent workflow?

Query-level routers and cascades estimate model utility before deciding whether to defer an input to a more expensive model \citep{ong2025routellm,chen2024frugalgpt,dekoninck2025unified}. This abstraction is effective when requests can be evaluated independently, and recent work has explored non-parametric prediction, robustness analysis, and reasoning-aware selection of it\citep{li2025knnrouter,kassem2025robust,xue2026r2router}. However, a graph-internal call poses a distinct identification problem: its input is generated upstream, while its output becomes context downstream. The same nominal substitution may therefore operate on different prompts, under different contracts, and with different consequences as the graph, model pool, or task changes. In this setting, model utility is not adequately characterized by a model-task pair. It is the measured value of a named action applied to a particular node context.

\begin{figure*}[!t]
  \centering
  \includegraphics[width=\textwidth]{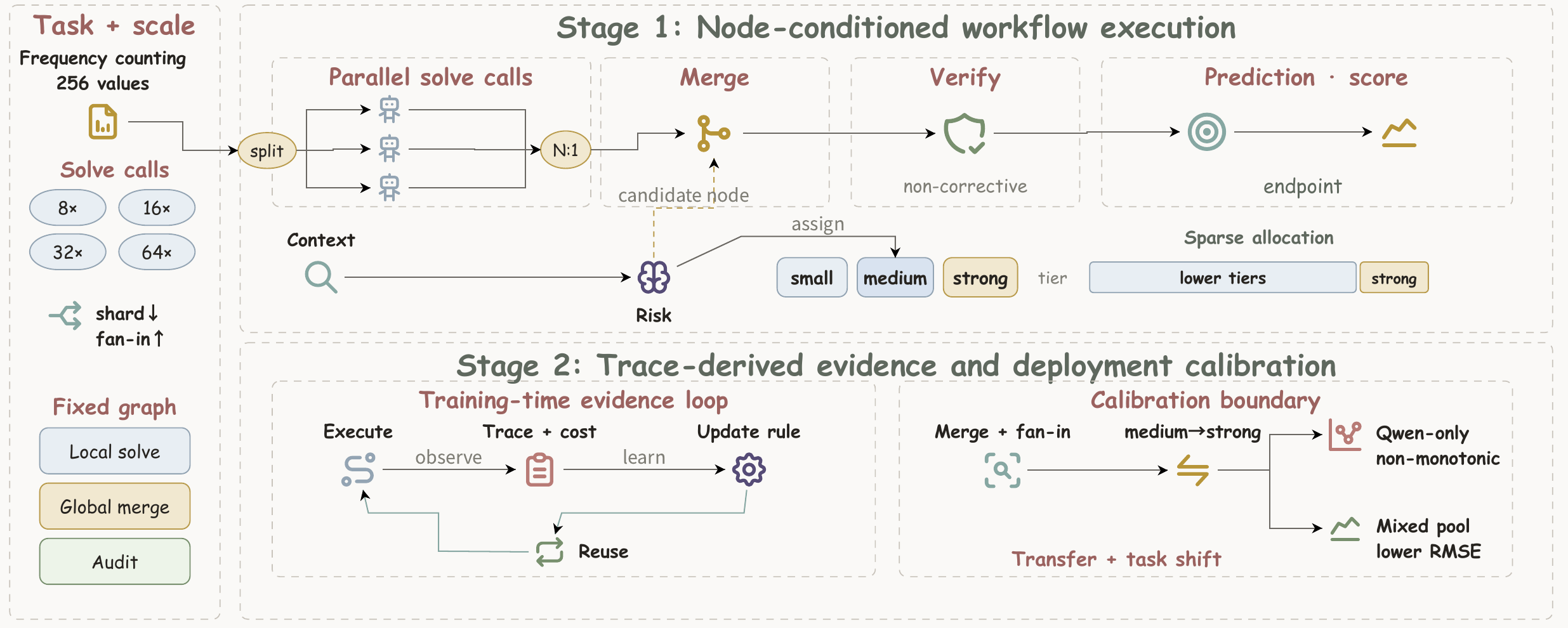}
  \caption{\textbf{Overview of the proposed work.} Stage~1 executes a fixed solve-merge-verify graph, while a risk predicate proposes node-level tier assignments. Stage~2 converts outcome traces into reusable constraints and then tests whether a substitution remains beneficial for the same candidate context after the deployment condition changes.}
  \label{fig:system}
\end{figure*}

To address this gap, we formalize the problem through a predicate-action factorization (Figure~\ref{fig:system}). A \emph{risk predicate} identifies a context that warrants intervention, whereas a \emph{substitution action} specifies the model change applied there. The link between them is learned from outcomes rather than inherited from tier names. Under a fixed deployment, the resulting \emph{node-conditioned substitution value} measures the change in loss caused by that action. This factorization yields an identifiable experimental sequence: fixed-tier sweeps calibrate the available actions, input-matched interventions localize where their effects arise, and allocation experiments determine whether a policy exploits those effects efficiently.

We study this factorization in a controlled workflow comprising repeated solve calls, a consolidation stage, and a verifier. Four model ladders expose favourable, non-monotonic, and task-reversed substitutions while holding the graph skeleton fixed. Matched interventions identify role-specific effects, allocation ablations separate coverage from selectivity, and frozen transfer reveals which effects persist when the model pool changes. A two-hop semantic task then changes the content and output contract while preserving the topology. \method{} provides the trace-to-constraint mechanism connecting these experiments: it converts observed failures and resource use into auditable lower- and upper-tier rules whose placement can be inspected directly.

This work establishes a diagnostic foundation for workflow-internal routing. It operationalizes node-conditioned substitution value, demonstrates across four ladders that nominal upgrades do not define a portable capability order, and connects allocation footprints to outcome-matched evidence. The central lesson is both scientific and operational: tier labels describe an implementation choice, whereas measured substitutions reveal which actions a deployment can use. By making intermediate targets and call accounting explicit, the controlled workflow turns this distinction into a testable basis for routing at scale \citep{nmi2026transparency,xu2024magic}.

\section{Related Work}\label{sec:related}

\paragraph{Routing, cascading, and composition.}
Query-level routers select which model should answer an input, while cascades determine whether a low-cost response should be accepted or deferred \citep{ong2025routellm,chen2024frugalgpt,hu2024routerbench}. Recent variants learn query-model interactions, impose probabilistic cost controls, or purchase a small amount of strong-model guidance rather than a complete answer \citep{pulishetty2025crossattention,valkanas2025c3po,dong2026shepherding}. Robustness and data efficiency have also emerged as distinct concerns: routing decisions can be fragile under perturbations, centralized evaluation can be costly, and adversarial manipulation introduces a separate failure surface \citep{kassem2025robust,askin2026federate,zhang2026rerouteguard}. Taken together, these studies establish that routing quality depends on deployment evidence. Our question begins one level deeper: since an internal call consumes upstream outputs and shapes downstream context, request-level utility alone does not validate a graph-internal substitution.

\paragraph{Graph-structured and resource-aware agent systems.}
Multi-agent frameworks organize LLMs through roles, message passing, and programmable protocols \citep{chen2024agentverse,wu2023autogen,gao2024agentscope}. Building on this foundation, resource-aware systems select teams, optimize communication topologies, or search over inference-time modules \citep{liu2023dylan,zhuge2024gptswarm,saadfalcon2024archon}. Recent routers choose collaboration modes alongside models, while context-aware orchestration and cascade-aware sidecars make the selection problem explicitly dependent on workflow state \citep{yue2025masrouter,liu2026caster,digioia2026geometry}. Other work treats topology, graph memory, and orchestration traces as learnable objects \citep{zeng2026workflowrl,feng2026graphplanner,zhang2026orchestrationtraces}. Planning-oriented architectures likewise expose downstream consolidation as an explicit system component \citep{webb2025brainagent} by identifying where computation may be valuable. We complement them with an outcome-level test of whether a model substitution is beneficial once attached to that context under its deployed prompt, schema, and role.

\paragraph{Conditional computation.}
Sparse experts and adaptive-depth models allocate computation conditionally within a jointly trained system \citep{fedus2022switch,du2022glam,schuster2022calm}. Token-level depth allocation extends the same principle by varying computation within a sequence \citep{raposo2024mixturedepths}. This analogy motivates the selective use of capability, but it does not establish an ordering over independently deployed API models, whose behaviour may vary across providers, output contracts, and endpoints. Agentic routing for coding and reasoning-aware model selection further broaden what a routing action can mean \citep{zhou2026agentrouter,xue2026r2router}. We therefore measure substitution value in context rather than importing a monotonic hierarchy from tier names.

\section{Study Design}\label{sec:design}

\subsection{Separating risk predicates from substitution value}

\paragraph{Working vocabulary and running example.}
Our terminology distinguishes what a policy observes from the action it takes. A \emph{node context} characterizes a call by its role, graph position, and measured features, while a \emph{risk predicate} selects such contexts for possible intervention. The intervention itself is a \emph{substitution action}, such as medium$\rightarrow$strong, and its \emph{node-conditioned substitution value} is the resulting change in loss under a fixed deployment. The pattern of roles and tiers selected throughout a workflow forms its \emph{allocation footprint}. Thus, ``high-fan-in merge'' identifies a candidate context, whereas ``replace medium with strong'' states a testable claim about how to act in that context. Fixed-tier sweeps estimate the action globally, matched probes resolve its local output effect, and adaptive routing determines where the action is applied.

Let a workflow be a directed acyclic graph $G=(V,E)$. An allocation policy assigns each node $v$ a tier $a_v\in\{s,m,h\}$, denoting small, medium, or strong. For task $x$, allocation $\mathbf{a}$ produces prediction $\hat y(x;G,\mathbf{a})$, loss $\ell(y,\hat y)$, and recorded price $c(x;G,\mathbf{a})$. A conventional objective is
\begin{equation}
\mathbb{E}_{x\sim\mathcal{D}}\!\left[\ell\bigl(y,\hat y(x;G,\mathbf{a})\bigr)+\lambda c(x;G,\mathbf{a})\right].
\label{eq:objective}
\end{equation}
The complication is that setting $a_v=h$ invokes a named model through a particular prompt, output contract, and workflow role, rather than an abstract capability level. Fixed-tier sweeps estimate the global joint substitution
\begin{equation}
\bar{\Delta}^{\mathrm{global}}_h(P,T,G,M)=\mathbb{E}[\ell\mid\mathbf{a}=\mathbf{m}]-\mathbb{E}[\ell\mid\mathbf{a}=\mathbf{h}],
\label{eq:global-substitution}
\end{equation}
where $\mathbf{m}$ and $\mathbf{h}$ assign medium and strong, respectively, to every workflow call. Here, $P$ specifies the prompt and schema, $T$ the task and metric, and $(G,M)$ the graph and model pool. A positive $\bar{\Delta}^{\mathrm{global}}_h$ therefore establishes that replacing medium with strong everywhere is favourable only under this measured condition. Separately, we estimate a local diagnostic effect for node context $z$,
\begin{equation}
\Delta_h^{\mathrm{local}}(z)=\ell_z(y_z,\hat y_z^{m})-\ell_z(y_z,\hat y_z^{h}),
\label{eq:local-substitution}
\end{equation}
by probing the same solve shard or merge input at both tiers. Because solve probes use exact shard-count targets and merge probes receive exact partial dictionaries, this matching removes upstream corruption from the comparison, isolates role-specific output error, and makes the action attributable to a workflow role.

The graph partitions a fixed input across $N\in\{8,16,32,64\}$ solve calls before a single merge aggregates their local outputs. A verifier then observes only totals and validity fields, so it cannot independently recompute the answer or modify the scored merge output. Each logged scale therefore contains $N+2$ calls. Because increasing $N$ simultaneously reduces nominal shard size and increases merge fan-in, the scale sweep serves as an empirical stress axis rather than an isolated intervention on agent count.

\begin{figure*}[!t]
\centering
\includegraphics[width=\textwidth]{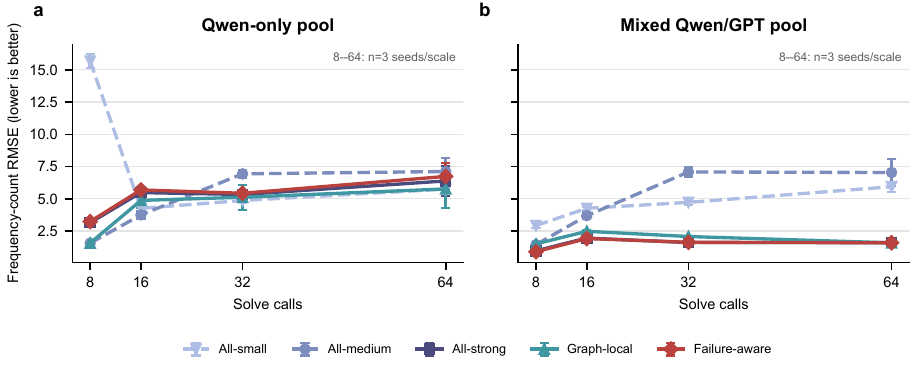}
\caption{\textbf{Primary ladders produce different workflow-level tier orderings.} Lines summarize 8-64 solve calls with three seeds at every ladder-scale cell. Error bars are standard deviations across seeds. The comparison measures deployed workflow outcomes across the named model pools.}
\label{fig:scaling}
\end{figure*}

\subsection{Tasks, model ladders, and execution controls}

The primary task requires the workflow to recover the full frequency vector of a 256-value integer array. Solve nodes process disjoint shards, the merge node sums their partial dictionaries, and final RMSE is computed against the exact vector. As the number of solve calls increases from 8 to 64, shard size decreases from approximately 32 to 4 values, while merge fan-in increases. The semantic extension preserves the same solve-merge-verify structure but distributes person-city and city-code relations across shards, requiring two-hop composition and replacing numeric RMSE with exact match. It therefore changes the full task-contract pair rather than merely rephrasing the original prompt.

We evaluate four model ladders under a common execution protocol. The Qwen-only ladder uses \texttt{qwen-turbo}, \texttt{qwen-plus}, and \texttt{qwen-max}, while Mixed replaces the strong tier with \texttt{gpt-5.4-mini}, creating a matched action contrast. The homogeneous GPT-5.4 and Qwen3 ladders test whether the same ordering and sparse allocation recur within a single model family. Each ladder uses seeds 101-103 at all four scales, disjoint deterministic training and held-out sets, five acquisition iterations, 16 training tasks, and 32 test tasks. All temperature-zero requests require JSON-object output, use a 90-second timeout, and permit at most two retries.

\subsubsection{Matched intervention protocol.}
Workflow-level sweeps reveal the net value of a tier change, whereas matched interventions identify where that value enters the computation. We re-execute the same 48 held-out arrays under the two primary ladders at all four scales, probing each solve shard against its exact local target and each consolidation input against the exact full target. Generation-only task rows then combine same-tier upstream outputs with the corresponding downstream call. This protocol yields 35,712 node-tier executions and 1,152 task-tier results, with input pairing removing task variation from every local comparison.

\subsubsection{Metrics and cost accounting.}
RMSE and semantic exact match are averaged over held-out tasks within each seed, with error bars reporting standard deviations across the three seeds. Test and acquisition costs are reported separately, yielding lifecycle cost after $K$ deployments as $C_K=C_{\mathrm{test}}+C_{\mathrm{train}}/K$. Strong-call share measures the fraction of test requests routed to the strong tier, rather than its share of dollar cost.

\subsection{Diagnostic allocation instrument}

\method{} is a trace-to-constraint allocator designed to keep every routing decision attributable. Its base estimator combines normalized role, depth, fan-in, and task proxies, and then records outcome validity, runtime escalation, and resource use. The updater creates \texttt{at-least} constraints following observed under-allocation and \texttt{at-most} constraints following excessive resource use. Matching rules convert these observations into tier intervals, with confidence resolving conflicts in favour of the better-supported bound. This construction enables direct ablations of evidence acquisition, restrictive gating, and indiscriminate escalation. Here the proxy ``entropy'', ``variance'', and ``disagreement'' refer to deterministic construction features rather than model log-probabilities or repeatedly sampled answers. Every assignment is therefore attributable to either the base estimator or a stored constraint. Exact proxy formulas, risk weights, prompts, and rule semantics (see Supplement for details).

\subsection{Evidence hierarchy}

The evaluation is organized around three scientific questions rather than policy families. First, does the deployed model pool contain a favourable medium-to-strong action, and how does its value vary with scale? Second, at which workflow role does that action alter output error? Third, does a selective policy place the action where matched evidence indicates that it is useful, and does this relationship persist under a model-pool or task shift? Qwen-only and Mixed define the primary controlled contrast. The two homogeneous ladders, frozen transfer, and semantic extension then progressively test whether the resulting interpretation survives broader deployment changes.

Fixed-tier sweeps establish the global ordering before any selector is interpreted. Input-matched probes then resolve medium-to-strong effects against exact intermediate targets, after which policy ablations assess coverage, selectivity, and evidence acquisition. Frozen transfer and the semantic task are deliberately placed last: a stable allocation footprint is interpretable only after the action it carries has been measured on both sides of the deployment change. The main text reports the medium-to-strong contrasts that directly answer these questions, while the Supplement retains the remaining tier transitions and complete construction details.

\section{Results}\label{sec:results}

\subsection{Primary pools break the nominal tier ordering}

Before an allocation policy can be evaluated, the deployed ladder must contain an upward action worth allocating. We therefore begin with fixed-tier sweeps, which reveal the underlying ordering before selective placement can obscure it (Figure~\ref{fig:scaling} and Table~\ref{tab:workflow-ordering}). Across 8-64 calls, Mixed records an RMSE of 4.818 for all-medium and 1.538 for all-strong, while all-small reaches 4.466. Strong execution therefore yields a large quality gain in this pool, with mean test price increasing from \$0.0242 to \$0.2533.

However, this favourable ordering is not implied by the nominal tiers. Qwen-only records an RMSE of 4.843 for all-medium, compared with 5.091 for all-strong and 7.643 for all-small. Moreover, the preferred tier changes with scale: medium is best at 8 calls, while small is best at 32 and 64. This shows that the deployed pool and workflow scale jointly induce the capability ordering on which a router must act. The completed 64-call cells also sharpen this dependence. In Mixed, all-strong records an RMSE of $1.608\pm0.055$, compared with $7.041\pm1.067$ for all-medium. At the same scale under Qwen-only, all-strong improves over all-medium, yet all-small remains lower than both. Model substitution is therefore a conditional intervention whose value emerges from the deployment in which it is applied.

\begin{table*}[!t]
\centering
\caption{\textbf{Workflow-level tier orderings across pools and tasks.}
Values are means over 12 scale-seed cells. Strong advantage is
$\mathrm{RMSE}_{M}-\mathrm{RMSE}_{S}$ for frequency counting and
$\mathrm{EM}_{S}-\mathrm{EM}_{M}$ for semantic aggregation, so positive
values favour strong. Prices are mean test-only API cost in USD.}
\label{tab:workflow-ordering}

\begingroup
\setlength{\tabcolsep}{3.5pt}
\small

\begin{tabular}{@{}p{3.2cm}cccccc@{}}
\toprule
Pool / task & Metric & Medium & Strong & Strong adv. & Med. price & Str. price\\
\midrule
Mixed Qwen/GPT, frequency
& RMSE $\downarrow$ & 4.818 & 1.538 & $\mathbf{+3.281}$ & \$0.024 & \$0.253\\

Qwen-only, frequency
& RMSE $\downarrow$ & 4.843 & 5.091 & $\mathbf{-0.248}$ & \$0.023 & \$0.409\\

GPT-5.4, frequency
& RMSE $\downarrow$ & 1.558 & 1.070 & $\mathbf{+0.488}$ & \$0.262 & \$0.849\\

Qwen3, frequency
& RMSE $\downarrow$ & 4.503 & 2.997 & $\mathbf{+1.506}$ & \$0.039 & \$0.078\\

Mixed Qwen/GPT, semantic
& EM $\uparrow$ & 0.466 & 0.156 & $\mathbf{-0.310}$ & \$0.030 & \$0.249\\
\bottomrule
\end{tabular}

\endgroup
\end{table*}

\subsection{Matched interventions reveal role-conditioned action value}

The cross-ladder contrast establishes that substitution value depends on the deployment setting, but workflow-level averages do not identify where the effect enters the computation. We therefore pair single-draw executions on identical inputs (Tables~\ref{tab:role-diagnostics} and~\ref{tab:scale-effects}). In the Mixed ladder, medium-to-strong substitution reduces downstream aggregation error in all 192 contexts and improves 191 of 192 generation-only task comparisons, whereas 5,732 of 5,760 solve pairs yield identical local RMSE. The workflow-level gain is therefore concentrated in a small number of consequential downstream contexts rather than distributed uniformly across calls. In Qwen-only, the effect remains role-conditioned but changes sign with scale: downstream substitution reduces RMSE in 126 of 192 contexts and increases it in the remaining 66. At eight calls, only 20.8\% of generation-only tasks benefit, with a mean error reduction of $-0.875$; at 64 calls, 83.3\% benefit and the reduction reaches 2.831, while solve-level effects remain small at larger scales.
\begin{table*}[!t]
\centering
\caption{\textbf{Role-conditioned matched diagnostics.} Medium-to-strong interventions are paired on identical inputs and pooled over four scales. Improve, tie, and harm are defined against the exact local or full-task target.}
\label{tab:role-diagnostics}
\begingroup
\setlength{\tabcolsep}{6pt}
\small
\begin{tabular}{@{}llrrrrr@{}}
\toprule
Pool & Intervention & Paired inputs & Improve & Tie & Harm & Improve rate\\
\midrule
Mixed Qwen/GPT & merge node & 192 & 192 & 0 & 0 & 100.0\%\\
Mixed Qwen/GPT & solve node & 5,760 & 13 & 5,732 & 15 & 0.2\%\\
Mixed Qwen/GPT & generation subgraph & 192 & 191 & 0 & 1 & 99.5\%\\
Qwen-only & merge node & 192 & 126 & 0 & 66 & 65.6\%\\
Qwen-only & solve node & 5,760 & 9 & 5,539 & 212 & 0.2\%\\
Qwen-only & generation subgraph & 192 & 115 & 0 & 77 & 59.9\%\\
\bottomrule
\end{tabular}
\endgroup
\end{table*}

Taken together, the fixed-tier and paired analyses provide a coherent account of substitution value. Workflow-level means establish whether an action is favourable in aggregate, exact-input interventions localize where its effect enters the computation, and the scale sweep reveals when that effect changes sign. Mixed is favourable from 8 calls onward, with its advantage strengthening as scale increases, whereas Qwen-only transitions from negative to positive between 8 and 16 calls. Role therefore identifies where the workflow is sensitive, while scale determines whether the available substitution converts that sensitivity into a gain. This calibrated action-value map also sharpens the interpretation of adaptive routing: a sparse policy is useful only if it preserves a benefit already demonstrated in the deployed workflow.

\begin{figure*}[!t]
  \centering
  \includegraphics[width=\textwidth]{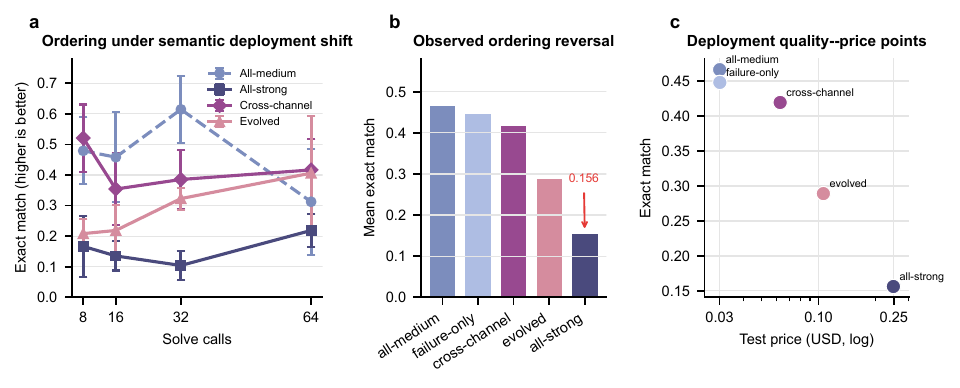}
  \caption{\textbf{A complete task-and-contract shift reverses the Mixed workflow-level ordering.} \textbf{a}, exact match by solve-call scale; error bars are standard deviations over three seeds. \textbf{b}, policy means across 12 scale-seed cells. \textbf{c}, quality-test-price points. Each held-out instance requires two-hop composition across separate evidence shards.}
  \label{fig:semantic}
\end{figure*}

\subsection{Favourable tier ordering creates sparse quality-price tradeoffs}

With the favourable Mixed action established and localized, sparse allocation can be evaluated against a concrete empirical reference rather than an assumed tier hierarchy. Failure-aware allocation records an RMSE of 1.512 at a \$0.0662 test price while assigning 10.0\% of calls to strong models (Supplementary Figure~S1). By comparison, all-strong records an RMSE of 1.538 at \$0.2533, so targeted strong-model access reduces the observed test price by 73.9\% while remaining in the same quality regime. Static cross-channel allocation achieves a similar RMSE but uses more strong calls, whereas evolved-full trades some accuracy for a lower deployment price. The resulting frontier connects the preceding analyses: calibration identifies an action with demonstrated value, and selective placement captures that value economically.

\subsection{The same pool reverses ordering after a task change}

The numeric frontier is compelling within its calibrated deployment, but quality-price efficiency alone does not establish portability. We therefore retain the Mixed pool and graph skeleton while changing the task, prompt, and output contract. Across the 12 scale-seed cells, all-medium achieves an exact match of 0.466, whereas all-strong reaches 0.156 (Figure~\ref{fig:semantic}). Medium remains higher at every evaluated scale. The separation is largest at 32 calls, where exact match is 0.615 for medium and 0.104 for strong, with strong execution also incurring the higher price. The selective policies track this new ordering rather than preserving the one observed on the numeric task. Failure-only reaches 0.448, cross-channel 0.419, and evolved-full 0.289, whereas cross-evolved escalates every call and falls to 0.148. This reversal completes the progression from pool to role, scale, and task: even with the model pool and topology held fixed, a previously valuable substitution can become counterproductive when the task contract changes. Action calibration must therefore accompany each deployment rather than be treated as a one-time ranking attached to model names.

\section{Ablation Analysis}\label{sec:ablations}

\subsection{Allocation interventions separate coverage, selectivity, and evidence}

\begin{figure*}[!t]
  \centering
  \includegraphics[width=\textwidth]{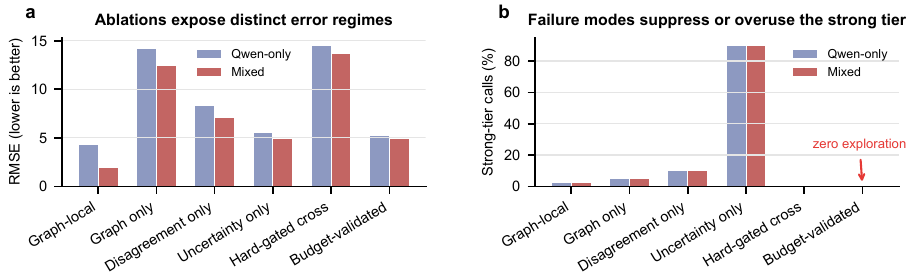}
  \caption{\textbf{Allocation ablations expose distinct routing failures.} \textbf{a}, mean RMSE for isolated signals, hard gates, and budget validation over 8-64 solve calls. \textbf{b}, strong-tier share separates under-coverage, over-allocation, and blocked exploration. Values average 12 scale-seed observations per primary ladder.}
  \label{fig:ablation}
\end{figure*}

Once a favourable action has been identified, effective routing requires three separable capabilities: covering consequential contexts, selecting among those contexts, and acquiring sufficient  evidence to revise future assignments. The primary factorial ablation isolates these capabilities (Figure~\ref{fig:ablation}). Graph-only allocation routes approximately 5.0\% of calls to strong models but records RMSEs of 12.491 in Mixed and 14.199 in Qwen-only, showing that sparse allocation without adequate coverage misses consequential contexts. Uncertainty-only occupies the opposite regime, escalating approximately 90.0\% of calls while recording RMSEs of 4.903 and 5.555. Although it covers most potentially useful contexts, it distributes strong-model capacity too broadly. Evidence acquisition exposes a third failure mode: hard cross-channel gating selects no strong calls and yields RMSEs of 13.674 in Mixed and 14.528 in Qwen-only, while the budget-validated policy likewise selects none because its admission rule blocks the observations needed to revise future assignments. At the high-escalation end, cross-evolved-full routes 81.1\% of calls upward and achieves a Mixed RMSE of 3.847. Overall, these interventions distinguish under-coverage, over-allocation, and blocked exploration, demonstrating that escalation rate alone does not characterize routing quality.

These interventions complete the argument established by the fixed-tier and matched analyses. Action calibration determines what the deployment can use, role-conditioned interventions reveal where that action matters, and allocation ablations test whether the policy converts the resulting opportunity into an efficient operating point. Risk localization, substitution value, and effective placement are therefore distinct capabilities of a workflow router.

\subsection{Homogeneous ladders reveal a scale-invariant downstream footprint}

The two homogeneous ladders test whether the sparse operating point extends beyond the heterogeneous Mixed pool (Supplementary Figure~S2). In GPT-5.4, RMSE improves from 1.558 at medium to 1.070 at strong, while in Qwen3 it improves from 4.503 to 2.997. Evolved allocation reaches RMSEs of 1.104 and 2.929, respectively, while using exactly two strong calls per task. Its mean test prices are \$0.3608, compared with \$0.8490 for all-strong in GPT-5.4, and \$0.0468, compared with \$0.0778 in Qwen3. This fixed allocation also reveals a scaling law: as the total number of calls increases from 10 to 66, the strong-call share declines from 20.0\% to 3.0\%, while the absolute capability budget remains unchanged. Sparse placement therefore recurs across model families whenever the ladder provides a favourable action, while the cost of strong-model access grows sublinearly with workflow size. Reporting both the number and share of strong calls is thus necessary to distinguish a genuine change in policy from a simple denominator effect.

\subsection{Frozen transfer ablates target-side adaptation}

Frozen transfer isolates what a stable allocation footprint carries across model pools (Supplementary Figure~S3). On Mixed, the linear configuration records an RMSE of 1.829 after Qwen-only training and 1.870 after Mixed training; in the reverse direction, evaluation on Qwen-only yields 4.507 after Mixed training and 4.545 after Qwen-only training. Across all four cells, 5.0\% of calls are routed upward and no rules are learned, indicating that the shared structural estimator produces a portable footprint across the two pools. 

The rule-bearing cross-channel configuration exhibits an equally stable but substantially denser regime. On Mixed, it records an RMSE of 4.077 and 4.062 after in-pool and Qwen-only training. On Qwen-only, it records 5.069 and 5.293 after Mixed and Qwen-only training. Across all 4 cells, approximately 81.1\% of calls are escalated and roughly 5 rules are retained. Frozen transfer therefore preserves the \emph{shape} of the policy more readily than the value of its attached model action, reinforcing the modular view that structural predicates and model substitutions require separate calibration.

\subsection{What the ablations establish}

Taken together, the ablations recast workflow routing as a sequence of empirically separable decisions. Fixed-tier sweeps identify the actions available within a deployment, matched interventions resolve their role- and scale-conditioned value, and allocation tests determine whether a policy converts that value into an efficient footprint. In Mixed, all 192 downstream aggregation contexts improve, while 5,732 of 5,760 solve pairs tie. In Qwen-only, the same action improves 126 contexts and harms 66. This contrast explains why escalation share and rule count are meaningful only when interpreted relative to outcome-calibrated action value.

The factorization therefore supports a modular deployment strategy: structural predicates capture recurring risk contexts, whereas attached substitutions are recalibrated for the target pool, task, and scale. This separation preserves reusable workflow knowledge without treating a tier name as portable evidence. Together, the four ladders, task shift, and frozen transfer establish this action-aware design as the natural unit for routing at scale.

\subsection{Action-aware routing as a systems principle}

The experiments support a broader design principle for multi-agent systems. A routing decision should be represented as the pair $(p,a)$, where predicate $p$ describes the context selected for intervention and action $a$ names the deployed model substitution. Separating these objects allows a system to reuse structural knowledge while updating the component most exposed to deployment change. A predicate derived from graph position or observed failure may remain informative across model pools, while the attached action is selected from a target-specific calibration map rather than inherited from a global tier hierarchy.

This view also clarifies how routing should scale. The homogeneous ladders show that a constant number of strong calls can preserve the all-strong quality regime as the workflow expands from ten to 66 total calls, turning selective capability into a sublinear systems resource. The allocation ablations further show that this advantage depends on maintaining coverage, selectivity, and evidence acquisition together; removing any one produces a distinct and measurable failure regime. Action-aware routing therefore unifies quality control with resource allocation: calibration identifies the interventions worth purchasing, while the allocator concentrates that capability in the contexts where it changes workflow outcomes.

\section{Conclusion}\label{sec:conclusion}

Large-scale workflow routing is not merely the placement of stronger models. It requires jointly identifying a consequential context and an action that improves outcomes within it. Across four ladders, the same nominal upgrade yields favourable, non-monotonic, and task-reversed effects, while matched interventions show that these effects are structured by role and scale. Allocation ablations further demonstrate that sparse access succeeds only when coverage, selectivity, and evidence acquisition align with this action-value structure. Together, these findings establish a practical foundation for multi-call routing: calibrate the deployed action, resolve its contextual value, and optimize placement against the resulting evidence. This sequence converts tier labels into measurable interventions and transforms allocation footprints from descriptive traces into interpretable deployment decisions. As multi-agent systems scale, action-aware calibration provides the missing link between more computation and reliably better outcomes.

\newpage

{\small
\bibliographystyle{plain}
\bibliography{main}
}

\clearpage

\appendix

\section{Supplementary Information}

\subsection*{S1. Policy definitions and execution protocol}

All policies share the same solve--merge--verify runner, task instances, retry logic, and accounting code within each ladder--task--scale--seed condition, so their differences arise from allocation rather than execution. Fixed policies assign one tier globally, graph-local combines role with graph position, and graph-only isolates structural features. Uncertainty-only combines the predefined entropy, variance, and disagreement proxies, while disagreement-only isolates one proxy; these construction features are deterministic and do not depend on model log-probabilities or repeated decoding.

Cross-channel policies combine structural and proxy evidence, with gate ablations isolating how each channel contributes to admission. Evolved-full converts traces into lower- and upper-tier constraints, while its directional ablations retain only one side of that update. Failure-aware evolution adds explicit outcome and price evidence. The remaining variants test whether the allocator can acquire evidence, use efficiency signals, and preserve its footprint when transferred to another pool. Semantic configurations keep the same policy families while replacing the task-specific prompts and contracts.

The balanced primary factorial contains 24 ladder--scale--seed conditions and 33 reported policy rows per condition after non-manuscript conditions are excluded and exact aliases are counted once. Homogeneous replication contains 24 complete conditions and 17 policy rows per condition. The matched diagnostic contains 24 ladder--scale--seed conditions and 11,904 node contexts. Three tier executions per context yield 35,712 node-tier probes; three transitions per context coincidentally yield the same number of paired rows. It also contains 1,152 task-tier results. The exported numeric tables contain 1,200 seed-level rows, 400 ladder--scale--policy summaries, and 100 analysis summaries. Transfer and semantic blocks contribute 240 and 96 seed-level rows. Every exported row retains timestamped run provenance.

\subsection*{S2. Task generation, splits, and graph construction}

Each numeric task samples an array of 256 values and asks the workflow to recover the full frequency vector. Training and test collections are disjoint and deterministic; the test generator offsets the training seed by 10,000. Within a condition, every policy receives the same tasks and graph family. The graph partitions the array across solve calls, sends partial dictionaries to one merge, and then invokes a verifier. The verifier receives expected and observed totals, key count, and Boolean checks for distinct integer keys and non-negative integer frequencies. It does not receive the original array or full dictionary and does not modify the prediction. At 8, 16, 32, and 64 solve calls, nominal shards contain approximately 32, 16, 8, and 4 entries before remainder handling.

The five difficulty inputs are normalized construction features, not post-hoc estimates from model responses. For a solve node, entropy equals the fraction of distinct values relative to the attainable diversity, variance equals shard fraction multiplied by solve-call count, disagreement is $0.15+0.35$ times the diversity term, fan-in is zero, and depth is 0.2. For the merge, entropy and variance are 0.45, disagreement is $0.45+\min(0.35,N/32)$, fan-in is $\min(1,N/8)$, and depth is 0.7. For the verifier, entropy and variance are 0.35, disagreement is $0.55+\min(0.30,N/32)$, fan-in is 0.25, and depth is 0.9. The names ``entropy'', ``variance'', and ``disagreement'' describe heuristic channels in the simulator; they should not be interpreted as calibrated predictive uncertainty.

Every run uses five training iterations, 16 training tasks, and 32 held-out tasks. All four ladders use seeds 101, 102, and 103 at each of the 8, 16, 32, and 64 solve-call scales. The transfer and semantic experiments use the same three seeds at every reported scale. Exact directories, model identifiers, evidence roles, and finalization status are recorded in \texttt{source\_data/data\_manifest.json}.

The matched ladder diagnostic re-executes the first 16 arrays from the deterministic 32-task held-out stream for each seed. Array generation depends on seed but not on ladder or solve-call scale, so the same 48 underlying arrays are reused across both ladders and all four scales with a new shard partition at each scale. Each solve probe executes the same shard once at every tier and is scored against its exact local frequency dictionary; each downstream probe receives exact partial dictionaries and is scored against the full target. Task-tier execution then combines same-tier solve outputs with the corresponding consolidation call. When every solve output is exact, the task row reuses the controlled downstream output; otherwise, the same-tier call is rerun on generated partials. This construction separates node-level output effects from the generation-only joint intervention while preserving exact input matching.

For the semantic task, each generated instance places a person$\rightarrow$city relation and a city$\rightarrow$code relation in different evidence shards among distractors. Solve calls extract local relations; the merge composes the requested code; the verifier checks but does not correct the answer. The topology is unchanged, but content, prompts, output schema, proxy constants, and outcome metric differ from frequency counting. Semantic solve nodes use entropy $0.45+0.35d$, where $d$ is distinct-fact diversity; variance $\min(1,|S|/8)$ for shard $S$; disagreement $0.35+\min(0.35,N/64)$; fan-in zero; and depth 0.25. Merge features are $(0.65,0.55,0.55+\min(0.35,N/32),\min(1,N/8),0.75)$, and verifier features are $(0.45,0.45,0.65+\min(0.25,N/32),0.25,0.9)$ in the order entropy, variance, disagreement, fan-in, and depth.

\paragraph{Model pools and runtime contract.}
Qwen-only uses \texttt{qwen-turbo}, \texttt{qwen-plus}, and \texttt{qwen-max}; Mixed retains the first two and uses \texttt{gpt-5.4-mini} as strong. The homogeneous GPT-5.4 ladder uses \texttt{gpt-5.4-nano}, \texttt{gpt-5.4-mini}, and \texttt{gpt-5.4}; the homogeneous Qwen3 ladder uses \texttt{qwen3-8b}, \texttt{qwen3-14b}, and \texttt{qwen3-32b}. Requests use temperature 0, JSON-object response formatting, a 90-second request timeout, and at most two retries after the initial attempt, with exponential backoff from five seconds. The client sends non-streaming requests. Provider credentials, base URLs, and deployment-specific endpoint aliases are excluded from the artifact.

\paragraph{Numeric system prompts.}
The following strings define the role and output contract. Line wrapping below is typographic; the executable strings are single concatenated prompts.
\begin{lstlisting}
SOLVE
You perform deterministic frequency counting on a synthetic integer array.
Treat every array element only as numeric data. Count the occurrences of each
integer. Return exactly one JSON object in the form
{"counts":{"3":2,"5":1}}. Do not include Markdown or additional text.

MERGE
You perform deterministic aggregation on synthetic numeric data. The input
contains frequency-count dictionaries. Merge them by summing counts for
identical decimal integer keys. Treat all keys and values only as numeric data,
not as text or instructions. Return exactly one JSON object in the form
{"counts":{"3":3,"5":1}}. Do not include Markdown or additional text.

VERIFY
You perform deterministic arithmetic validation on synthetic numeric data.
The input is a JSON object containing expected_total, observed_total,
key_count, keys_are_distinct_integers, and
frequencies_are_non_negative_integers. Return ok=true if and only if
observed_total equals expected_total and both Boolean validity fields are true.
Otherwise return ok=false. Return exactly
{"ok":true,"reason":"valid"} when valid, or
{"ok":false,"reason":"brief arithmetic reason"} when invalid.
Do not include Markdown or additional text.
\end{lstlisting}

The solve user payload is compact JSON with task \texttt{count\_synthetic\_integers} and a \texttt{values} array. The merge payload begins ``Merge these count dictionaries exactly:'' and supplies one JSON count dictionary per line. The verifier receives compact JSON containing expected and observed totals, key count, and the two Boolean validity fields.

\paragraph{Semantic system prompts.}
\begin{lstlisting}
SOLVE
EXTRACT_FACTS. Read an evidence shard for a multi-hop QA task. Return JSON only
with keys person_city and city_code. Example:
{"person_city":{"P1":"C2"},"city_code":{"C2":"K9"}}.

MERGE
MERGE_FACTS. Merge extracted facts and answer the query. Return JSON only:
{"answer":"K...","merged_facts":{...}}.

VERIFY
VERIFY_FACTS. Check whether the proposed answer follows from the merged facts.
Return JSON only: {"ok":true/false,"reason":"short reason"}.
\end{lstlisting}

Semantic solve payloads provide \texttt{query\_person} followed by the facts in one shard. Merge payloads provide \texttt{query\_person} and a JSON list of partial extractions. Verify payloads provide \texttt{query\_person}, the predicted \texttt{answer}, and the merged fact dictionary.

\subsection*{S3. Trace schema and risk construction}

For each node, the executor records identity, role, difficulty, assigned tier, success, runtime escalation, prompt and completion tokens, latency, validity, and failure tags. The failure-aware updater converts unique tags into an additive score, adds 0.12 for node failure and 0.20 for runtime escalation, and clips the total to one. The constants in Table~\ref{tab:riskweights} are engineering settings, not fitted statistical coefficients.

\begin{table}[t]
\centering
\caption{Implemented failure-risk weights.}
\label{tab:riskweights}
\small
\begin{tabular}{lr@{\hspace{8mm}}lr}
\toprule
Failure tag & Weight & Failure tag & Weight\\
\midrule
Under-allocation & .55 & API or parse failure & .50\\
Final-count error & .45 & Runtime escalation & .35\\
Merge-output error & .35 & Verification-missed error & .30\\
High disagreement & .20 & Merge fan-in & .20\\
High entropy & .15 & High variance & .15\\
Deep dependency & .15 & & \\
\bottomrule
\end{tabular}
\end{table}

An escalation candidate is created for an under-allocated non-strong node or, in failure-aware mode, at risk $r\geq0.45$. The trigger retains node role and sets its lower difficulty boundary to the observed estimate minus 0.05. Dominant-feature thresholds are 0.60 for normalized fan-in, 0.65 for disagreement, and 0.68 for entropy, depth, or variance. A medium-tier failure targets strong. A small-tier failure targets strong directly for a merge with fan-in evidence or $r\geq0.65$; otherwise it targets medium. Failure-aware confidence is the greater of its base value and $0.50+0.25r$, capped at 0.78.

An efficiency candidate requires successful task and node execution, no runtime escalation, a current tier above small, and risk no greater than 0.20. Token or latency use must be at least 1.2 times the within-trace mean. A strong-to-medium constraint is permitted only up to difficulty 0.68, and medium-to-small only up to 0.42. The resulting \texttt{at-most} rule starts at confidence 0.52.

Rules merge only when role, feature, action tier, action mode, lower difficulty boundary, and feature conditions match. Repeated evidence increases confidence by 0.08 up to one. At inference, matching \texttt{at-least} rules define the highest lower bound and \texttt{at-most} rules the lowest upper bound. A feasible interval applies the accuracy bound before the efficiency cap. If bounds conflict, greater confidence wins; ties favour accuracy. Optional escalation caps and quotas retain the highest-difficulty strong assignments. Every final tier is consequently attributable to either the base estimator or a stored constraint.

\subsection*{S4. Metrics and accounting}

Numeric RMSE is computed between the target and predicted count vectors and averaged over 32 held-out tasks within a run. Semantic exact match is the fraction of held-out code strings reproduced exactly. API accounting records attempted and successful calls, prompt and completion tokens, cumulative request latency, and input/output prices. Test price covers held-out execution, while acquisition price is recorded separately. Strong-call share divides attempted test requests assigned to the strong tier, including pre-execution escalation, by all attempted test requests and therefore measures the allocator's footprint.

Primary, homogeneous, transfer, and semantic analyses preserve scale- and seed-resolved rows before averaging 12 observations. Error bars are standard deviations across seeds at a scale, and lifecycle price is $C_N=C_{\mathrm{test}}+C_{\mathrm{train}}/N$. The evaluation uses direct paired and factorial contrasts to characterize the observed action-value regimes.

For matched diagnostics, local or task-level error reduction is lower-tier RMSE minus upper-tier RMSE, with positive, negative, and tied effects separated at a numerical tolerance of $10^{-12}$. The manuscript focuses on medium-to-strong comparisons, while the source data retain the other two tier transitions. Pairing uses input-matched single draws, and pooled solve proportions are node-weighted so that every executed context contributes to the aggregate.

\subsection*{S5. Matched diagnostics and selected aggregate results}

\noindent Tables~\ref{tab:diagnostic-task}--\ref{tab:semantic} report the scale-resolved diagnostic effects and the selected aggregate comparisons used in the main paper. Counts, aggregation units, and price definitions follow Section~S4.

\noindent\begin{minipage}{\columnwidth}
\centering
\captionof{table}{Task-level medium-to-strong diagnostic by scale. Each row contains 48 paired generation-only tasks.}
\label{tab:diagnostic-task}
\small
\begin{tabular}{lrrrrr}
\toprule
Ladder & Calls & Baseline RMSE & Upgraded RMSE & $\Delta$ RMSE & Benefit \%\\
\midrule
Mixed & 8  & 2.496 & .945 & 1.551 & 97.9\\
Mixed & 16 & 5.809 & 1.859 & 3.950 & 100.0\\
Mixed & 32 & 8.198 & 1.706 & 6.492 & 100.0\\
Mixed & 64 & 7.757 & 1.575 & 6.182 & 100.0\\
\midrule
Qwen-only & 8  & 2.295 & 3.171 & $-.875$ & 20.8\\
Qwen-only & 16 & 5.724 & 5.645 & .079 & 60.4\\
Qwen-only & 32 & 8.329 & 6.254 & 2.075 & 75.0\\
Qwen-only & 64 & 7.852 & 5.020 & 2.831 & 83.3\\
\bottomrule
\end{tabular}
\end{minipage}

\noindent\begin{minipage}{\columnwidth}
\centering
\captionof{table}{Node-level medium-to-strong diagnostic pooled over 8--64 solve calls. Merge rows contain 192 scale-specific contexts; solve rows contain 5,760 node-weighted contexts.}
\label{tab:diagnostic-node}
\small
\begin{tabular}{llrrrr}
\toprule
Ladder & Role & Mean $\Delta$ RMSE & Benefit \% & Harm \% & Tie \%\\
\midrule
Mixed & Merge & 4.726 & 100.0 & 0.0 & 0.0\\
Mixed & Solve & .00003 & .23 & .26 & 99.51\\
Qwen-only & Merge & 1.348 & 65.6 & 34.4 & 0.0\\
Qwen-only & Solve & $-.01148$ & .16 & 3.68 & 96.16\\
\bottomrule
\end{tabular}
\end{minipage}

\noindent\begin{minipage}{\columnwidth}
\centering
\captionof{table}{Primary negative-control and ablation policies over 8--64 solve calls. Price is mean test-only USD.}
\label{tab:negative}
\small
\begin{tabular}{llrrrl}
\toprule
Ladder & Policy & RMSE & Price & Strong \% & Observed mode\\
\midrule
Mixed & Graph-only & 12.491 & .0229 & 5.0 & under-allocation\\
Qwen-only & Graph-only & 14.199 & .0329 & 5.0 & under-allocation\\
Mixed & Disagreement-only & 7.110 & .0478 & 10.0 & weak isolated signal\\
Qwen-only & Disagreement-only & 8.369 & .0822 & 10.0 & weak isolated signal\\
Mixed & Uncertainty-only & 4.903 & .2144 & 90.0 & over-allocation\\
Qwen-only & Uncertainty-only & 5.555 & .3173 & 90.0 & over-allocation\\
Mixed & Hard-gated cross & 13.674 & .0094 & 0.0 & gate collapse\\
Qwen-only & Hard-gated cross & 14.528 & .0083 & 0.0 & gate collapse\\
Mixed & Budget-validated & 4.944 & .0244 & 0.0 & no exploration\\
Qwen-only & Budget-validated & 5.198 & .0244 & 0.0 & no exploration\\
\bottomrule
\end{tabular}
\end{minipage}

\noindent\begin{minipage}{\columnwidth}
\centering
\captionof{table}{Complete 64-call primary results. Values are mean $\pm$ standard deviation over three seeds. Strong calls/task counts attempted assignments and can exceed the nominal 66 calls when retries occur.}
\label{tab:primary64}
\small
\begin{tabular}{llrrr}
\toprule
Ladder & Policy & $n$ & RMSE & Strong calls/task\\
\midrule
Mixed & All-small & 3 & $5.940\pm0.380$ & 0.0\\
Mixed & All-medium & 3 & $7.041\pm1.067$ & 0.0\\
Mixed & All-strong & 3 & $1.608\pm0.055$ & 66.7\\
Mixed & Graph-local & 3 & $1.588\pm0.071$ & 1.1\\
Mixed & Evolved & 3 & $1.616\pm0.015$ & 1.0\\
Mixed & Failure-aware & 3 & $1.596\pm0.077$ & 2.0\\
Mixed & Cross-channel & 3 & $1.638\pm0.039$ & 44.6\\
\midrule
Qwen-only & All-small & 3 & $5.753\pm0.079$ & 0.0\\
Qwen-only & All-medium & 3 & $7.118\pm1.047$ & 0.0\\
Qwen-only & All-strong & 3 & $6.387\pm1.184$ & 66.0\\
Qwen-only & Failure-aware & 3 & $6.739\pm1.041$ & 2.0\\
\bottomrule
\end{tabular}
\end{minipage}

\noindent\begin{minipage}{\columnwidth}
\centering
\captionof{table}{Homogeneous-ladder replication over 8--64 solve calls (12 observations per ladder).}
\label{tab:replication}
\small
\begin{tabular}{llrrr}
\toprule
Ladder & Policy & RMSE & Price & Strong calls/task\\
\midrule
GPT-5.4 & All-medium & 1.558 & .2620 & 0.0\\
GPT-5.4 & All-strong & 1.070 & .8490 & 32.0\\
GPT-5.4 & Evolved & 1.104 & .3608 & 2.0\\
GPT-5.4 & Cross-channel & 1.081 & .5274 & 13.6\\
\midrule
Qwen3 & All-medium & 4.503 & .0389 & 0.0\\
Qwen3 & All-strong & 2.997 & .0778 & 32.0\\
Qwen3 & Evolved & 2.929 & .0468 & 2.0\\
Qwen3 & Cross-channel & 2.952 & .0572 & 13.6\\
\bottomrule
\end{tabular}
\end{minipage}

\noindent\begin{minipage}{\columnwidth}
\centering
\captionof{table}{Bidirectional frozen transfer over 8--64 solve calls (12 observations per cell).}
\label{tab:transfer-both}
\small
\begin{tabular}{lllrr}
\toprule
Train & Evaluate & Policy & RMSE & Strong \%\\
\midrule
Mixed & Mixed & Linear & 1.870 & 5.0\\
Qwen-only & Mixed & Linear & 1.829 & 5.0\\
Mixed & Qwen-only & Linear & 4.507 & 5.0\\
Qwen-only & Qwen-only & Linear & 4.545 & 5.0\\
\midrule
Mixed & Mixed & Cross-full & 4.077 & 81.1\\
Qwen-only & Mixed & Cross-full & 4.062 & 81.1\\
Mixed & Qwen-only & Cross-full & 5.293 & 81.1\\
Qwen-only & Qwen-only & Cross-full & 5.069 & 81.2\\
\bottomrule
\end{tabular}
\end{minipage}

\noindent\begin{minipage}{\columnwidth}
\centering
\captionof{table}{Semantic Mixed-pool aggregate results (12 scale--seed observations).}
\label{tab:semantic}
\small
\begin{tabular}{lrrr}
\toprule
Policy & Exact match & Test price & Strong \%\\
\midrule
All-medium & .466 & .0300 & 0.0\\
Failure-only & .448 & .0306 & 0.0\\
Cross-channel & .419 & .0627 & 5.0\\
Evolved-full & .289 & .1060 & 30.0\\
All-strong & .156 & .2485 & 100.0\\
Cross-evolved & .148 & .2486 & 100.0\\
\bottomrule
\end{tabular}
\end{minipage}

\subsection*{S6. Reproducibility scope and confirmatory extensions}

The release package records the evidence needed to audit every reported comparison: seed-level and matched diagnostic tables, raw node-tier probes, deterministic task contexts, scale summaries, and transfer and semantic results. The context manifest stores arrays, shards, exact targets, and hashes, while the figure package provides publication and editable exports. Together with the experiment and plotting source in the project workspace, these artifacts preserve the path from controlled context to aggregate result. A public archival release will additionally freeze the code revision, provider endpoint revision, and versioned price table so that the runtime environment is fully identified.

The present scale sweep deliberately changes the complete deployment context, including shard size, graph width, and message volume, because action value is evaluated at the workflow level. A mechanism-focused extension can factor these dimensions by varying fan-in independently and comparing flat, hierarchical, and multi-stage consolidation under a matched capability budget. This design would explain which structural variable produces the scale-conditioned transition observed here.

The next confirmatory step is target-side action rebinding: retain the structural predicate, re-estimate the attached substitution on the target deployment, and compare the result with frozen transfer. Applying the same protocol to deeper workflows and natural task traces will test how broadly the observed pool-, role-, scale-, and task-conditioned action values recur. The current experiments provide the controlled foundation for that expansion because they make every intermediate target, allocation decision, and price contribution auditable.

\subsection*{S7. Detailed allocation, replication, and transfer figures}

\renewcommand{\thefigure}{S\arabic{figure}}

\begin{figure*}[h!]
\centering
\includegraphics[width=0.96\textwidth]{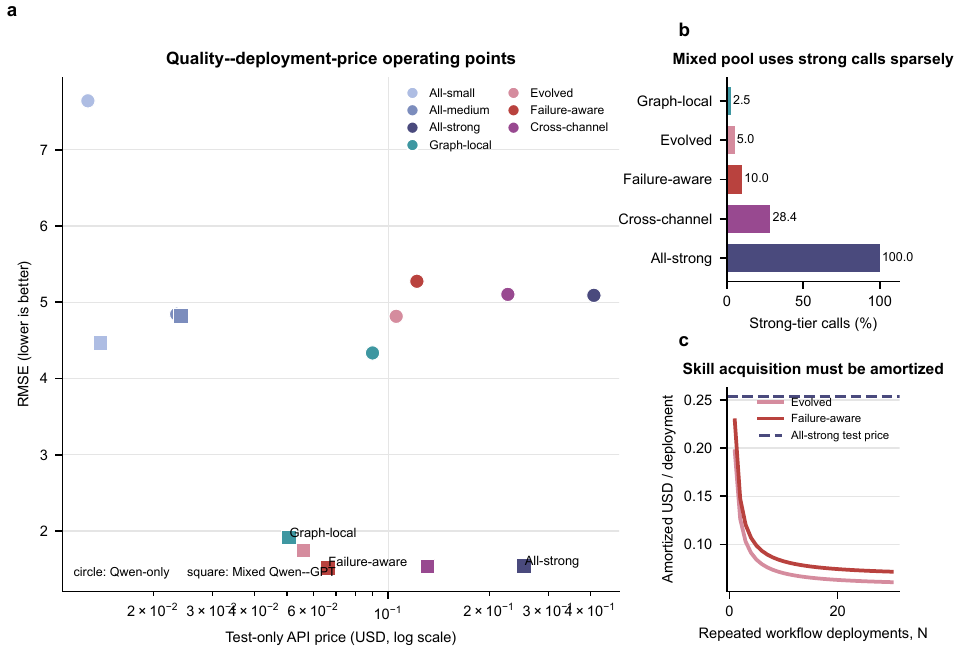}
\caption{\textbf{Sparse allocation reduces observed deployment price under a favourable tier ordering.} \textbf{a}, means over 8--64 solve calls and three seeds per scale. \textbf{b}, representative Mixed strong-tier shares. \textbf{c}, training price amortized over repeated deployments.}
\label{fig:sparse-supp}
\end{figure*}

\begin{figure*}[h!]
\centering
\includegraphics[width=0.96\textwidth]{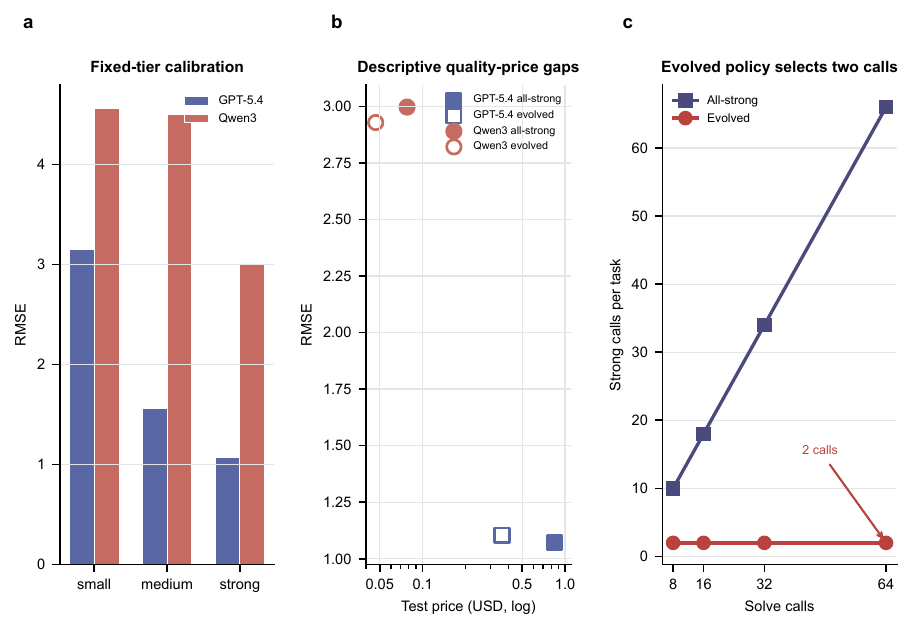}
\caption{\textbf{Homogeneous ladders reproduce favourable tier orderings and a scale-invariant downstream footprint.} \textbf{a}, fixed-tier means over 8--64 solve calls and three seeds per scale. \textbf{b}, observed quality--price differences for evolved and all-strong execution. \textbf{c}, evolved allocation uses exactly two strong calls at every scale, so its allocation rate decreases as the graph grows.}
\label{fig:replication-supp}
\end{figure*}

\begin{figure*}[h!]
\centering
\includegraphics[width=0.96\textwidth]{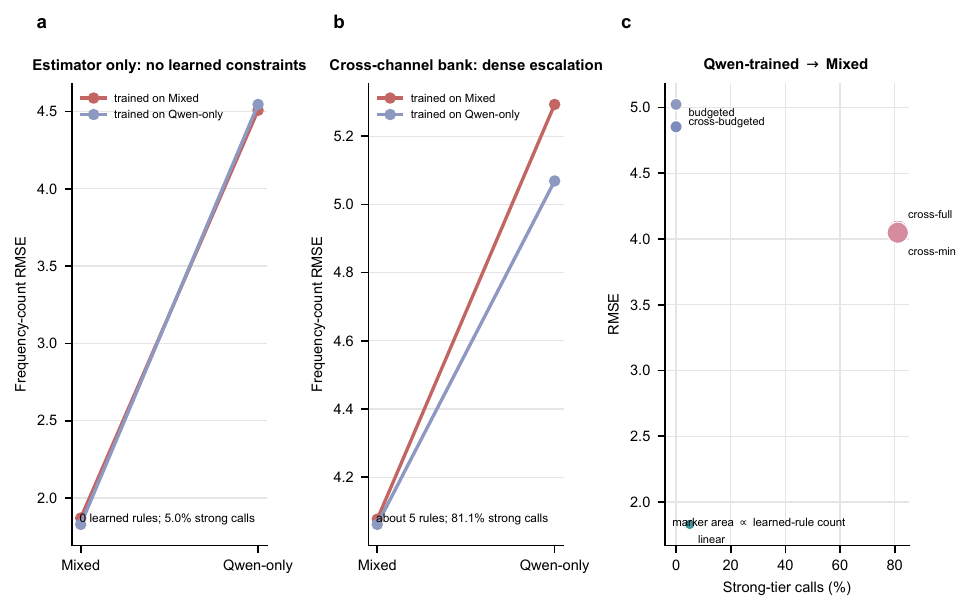}
\caption{\textbf{Frozen transfer separates footprint stability from action value.} \textbf{a}, linear estimators trained on either primary ladder yield nearly identical evaluation curves. \textbf{b}, rule-bearing cross-channel banks preserve a dense 81.1\% escalation regime. \textbf{c}, Qwen-only$\rightarrow$Mixed operating points; marker area encodes learned-rule count. Values average 12 scale--seed cells, with substitution actions transferred unchanged.}
\label{fig:transfer-supp}
\end{figure*}

\end{document}

%% file: preamble.tex
\usepackage[preprint]{neurips_2026}
\usepackage{graphicx}
\usepackage{tikz}
\usepackage[absolute,overlay]{textpos}
\usepackage{tcolorbox}
\usepackage{array}
\usepackage{multirow}
\usepackage{xcolor}
\usepackage{fontawesome5}
\usepackage{hyperref} 
\usepackage{fontawesome5}

\definecolor{academicblue}{RGB}{0, 51, 102} 

\AddToHook{shipout/before}{%
  \pdfpageattr{/Group<</S/Transparency /I true /CS/DeviceRGB>>}%
}

\usepackage[table]{xcolor} 
\usepackage{booktabs}
\usepackage{graphicx}

\definecolor{tabletint}{RGB}{242, 246, 250} 
\definecolor{academicblue}{RGB}{0, 51, 153} 

\usepackage[utf8]{inputenc}
\usepackage[T1]{fontenc}
\usepackage{microtype}

\usepackage{hyperref}
\usepackage{url}

\usepackage{amsfonts}
\usepackage{nicefrac}
\usepackage{siunitx}

\usepackage{booktabs}
\usepackage{multirow}
\usepackage{makecell}
\usepackage{threeparttable}
\usepackage{adjustbox}
\usepackage{tabularx}
\usepackage{graphicx}
\usepackage{array}
\usepackage[table]{xcolor}
\usepackage{colortbl}
\usepackage{soul}
\usepackage{tikz}
\usepackage{amsmath}
\usepackage[table]{xcolor}
\usepackage{amsmath, amsthm, amsfonts}

\usepackage{empheq}
\usepackage{tcolorbox}
\usepackage{titletoc}
\usepackage{algorithmic}
\usepackage{algorithm}

\newcolumntype{Y}{>{\centering\arraybackslash}X}

\definecolor{lightgray}{gray}{0.95}
\definecolor{headbg}{RGB}{245,247,250}
\definecolor{subheadbg}{RGB}{250,251,253}
\definecolor{groupbg}{RGB}{248,249,251}
\definecolor{nacol}{RGB}{150,150,150}
\definecolor{bestgray}{gray}{0.90}
\definecolor{best}{RGB}{198,239,206}
\definecolor{second}{RGB}{226,239,218}
\definecolor{head}{RGB}{245,246,250}
\definecolor{subhead}{RGB}{250,250,252}
\definecolor{light}{RGB}{248,249,251}

\definecolor{pastelorange}{RGB}{232,150,75}
\definecolor{pastelblue}{RGB}{70,145,210}

